\documentclass[prl,twocolumn,showpacs,superscriptaddress]{revtex4}
\usepackage{graphicx}
\usepackage{dcolumn}

\usepackage{amssymb}

\begin{document}
\title{Effect of secondary electron emission on subnanosecond breakdown 
 in high-voltage pulse discharge}
\author{I V Schweigert$^1$, A L Alexandrov$^1$,
 P Gugin$^2$, M Lavrukhin$^2$, P A Bokhan$^2$, Dm E Zakrevsky$^2$}
\affiliation{$^1$ Khristianovich Institute of Theoretical and Applied Mechanics, 
Novosibirsk 630090, Russia}
\affiliation{$^2$ A V Rzhanov Institute of Semiconductor Physics, 
 Novosibirsk 630090, Russia}

\date{\today }

\begin{abstract}

 A subnanosecond breakdown in high-voltage pulse discharge 
  is studied in experiment and 
 in kinetic simulations for mid-high pressure in helium.
 It is shown that the characteristic time of
 the current growth can be controlled by the secondary electron emission.
 We test the influence of secondary electron yield on plasma parameters
 for three types of cathodes  made from titanium, silicon carbide 
 and CuAlMg-alloy. By changing the pulse voltage amplitude and 
 gas pressure, the area of existence of subnanosecond breakdown
 is identified. 

\end{abstract}

\maketitle

\section {Introduction}

 Recently serious attention is paid to the study of physical 
 phenomena of subnanosecond current development in 
 discharge plasma in super-high-electric fields at 
 mid- and high-pressures. 
 Practical interest to this type of discharges 
is related to promising prospects for the development of novel 
electrophysical devices \cite{1,2}. For example, low- and high-energy 
electron beam 
sources can be applied for laser excitation or modification of 
materials \cite{3,4,5,6}. They also can be used as radiation sources
 \cite{7,8}, 
including x-ray radiation \cite{2}, high-pressure 
discharge pre-ionization devices \cite{5}, 
high-voltage subnanosecond \cite{1,2}
and picosecond pulse generation and commutation devices
 \cite{9}.  
 Nevertheless a complete model 
of discharge ignition and operation in super-high-electric fields 
  (E/N$>$10$^4$~Td) is still missing because of
 the lack of understanding of physics and elementary 
 processes proceeding under 
 high field conditions on subnanosecond time scale.

 Experimental study of breakdown in high voltage pulse (HVP)  
discharges in helium were done in Refs.\cite{9,10}. 
 There the current exponential growth within a subnanosecond  
 was registered in plasma sustained  
 between two plane cathodes and an mesh-anode between them. 
Applying 20~kV voltage,  the electron beams 
from the cathodes with the current density of 200~A/cm$^2$
 were generated. 
 The current growth rate was 500~A/cm$^2$ns. 

The kinetic model of this HVP discharge was developed
 in Refs. \cite{Sch2014,Sch2015,Sch2016}.                    
 The  model 
  includes the inelastic and elastic scattering of electrons, 
ions and energetic atoms with background helium, as well as  
 electron emission due to the photoemission and the bombardment of 
 the cathode surface by electrons, ions and energetic atoms.
The kinetic simulations with Particle-in-cell Monte Carlo collision
 (PIC MCC) method revealed unexpected scenario of the current 
 development in the HVP discharge.  
Before the electron emission 
due to ions and energetic atoms bombardment 
 was usually considered 
 as a main process in developing the breakdown 
 in abnormal glow discharge. 
However the subnanosecond breakdown in the HVP discharge
can not be supported by only electron emission with heavy particle
 bombardment due to their inertia. The electron emission 
 by photons radiated by atoms excited by electron impacts is also unable 
 to maintain the current growth rate observed in the experiments. This
 type of resonant photons is trapped in plasma due to reabsorption.
 Thus some additional mechanism of electron emission was required to
 describe the subnanosecond breakdown.  

 In Refs. \cite{9,Sch2014}, two additional processes 
 participating in electron emission from the cathode 
 such as the photoemission 
 by photons with Doppler shifted (DS) frequency 
 and secondary electron emission (SEE) have been proposed. 
The photons with DS frequency are 
 produced in discharge plasma 
 in the excitation reactions between the background atoms and 
 ions or energetic atoms as well as
  in the excitation transfer reactions between 
  excited atoms and energetic atoms. A feature of 
the DS photons is that they cross the 
discharge plasma without reabsorption. In contrast, 
 the photons from the impact excitation by electrons remain in the plasma due to multiple 
reabsorption for the time much larger than the breakdown one. 
In PIC MCC simulations \cite{Sch2014}, 
 the DS photons were shown to be very important during the initial stage of 
breakdown. The SEE becomes a major process in 
current growth at the final stage of breakdown when
 with decreasing voltage   
 the electrons accumulated in plasma during discharge operation 
 sweep over the cathodes. 
\begin{figure}[h!]
\includegraphics[width=0.8\linewidth]{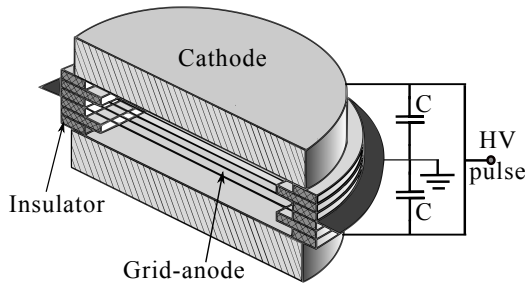}
\caption{Scheme of planar discharge cell with 
generation of two electron beams oppositely directed.}
\label{setup}
\end{figure}
 In this paper, in the experiment and in PIC MCC 
simulations we study the breakdown development in the HVP discharge 
 with three types of cathodes made from different 
 materials.
 All these materials have enhanced secondary electron emission yield. 
 Our purpose is to find a way to decrease the discharge breakdown 
 time by testing different cathode materials and changing 
 the gas pressure and voltage.

\section{Experimental setup}

The breakdown in the  high-voltage pulse
 discharge in helium is studied in the experimental cell shown 
in Fig. \ref{setup}. 
Two round cathodes with the total area of 1.6 cm$^2$ 
are placed 6 mm apart. A mesh-anode with the geometric 
transparency of 0.7 is placed between the cathodes. 
All electrodes are isolated with a set of plates made from 
 glass. 
 The pulse voltage is simultaneously applied to both cathodes and
  two oppositely directed electron beams are generated  due to cathode 
 emission.  The voltage amplitude $U_0$ ranges from  4~kV to 12~kV. 
 The gas pressure varies from 10~Torr to 35~Torr.            
 The cathodes are symmetrically connected to the external
low-inductance circuit and the mesh-anode is grounded.  
The pulse shape is registered with the low-inductive resistive 
 divider with the rate about 20:1 using oscilloscope Tektronix DPO 70804C 
 with a bandwidth of 8 GHz. 
The registration circuit and other experimental details were described 
 in Ref.\cite{exp_d}.

\begin{figure}[h!]
\includegraphics[width=0.85\linewidth]{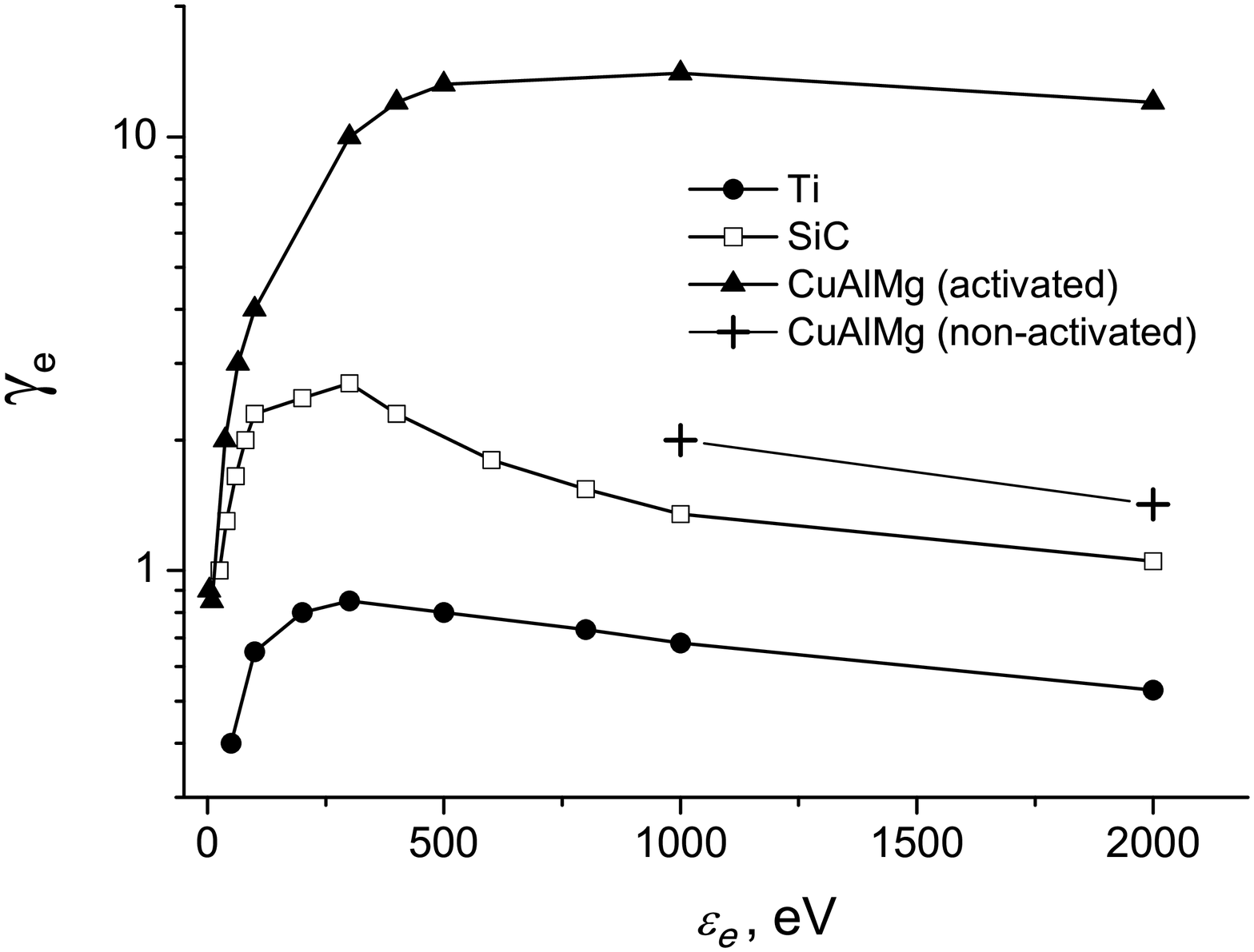}
\caption{Secondary electron emission coefficient as function of 
 the electron energy for cathodes made from titanium (full circle), 
 silicon carbide (open squares),  activated CuAlMg-alloy 
(triangles) and non-activated CuAlMg-alloy (+).}
\label{see}
\end{figure}

 In the experiments, the cathodes made from 
 titanium (Ti), silicon carbide (SiC), and CuAlMg alloy
  were tested.  All these materials  have large
 SEE coefficient  $\gamma_e$, 
 but the dependence of $\gamma_e$ from the electron 
 energy $\epsilon_e$ is very different. 
 The $\gamma_e$ as a function of an electron energy $\epsilon_e$
 is shown in Fig. \ref{see} for three types of 
 cathode materials.
 For titanium the $\gamma_e$ varies from 0.4 to 0.9 
 in the range of $\epsilon_e$ from 50~eV to 2~keV, with 
 maximum at 300 eV  \cite{Ti}.  
 For SiC the $\gamma_e$ has similar behavior  
  for $\epsilon_e$=20~eV-400~eV \cite{SiC}, 
 but it is nearly three times larger, $\gamma_e$= 0.9- 2.5. 
 The $\gamma_e$ for CuAlMg alloy can be very different depending
 on the treatment procedure and can be in activated or non-activated 
 states \cite{Lep}. 
  The activation take places due to 
 diffusion of Mg to the surface and its oxidation during annealing. 
 So, the activated  CuAlMg-alloy demonstrates  $\gamma_e$  up to 12.5 within the electron 
 energy range from 500~eV to 2~keV. 
For non-activated CuAlMg alloy the $\gamma_e$ is close to
 the SiC-$\gamma_e$.

The comparison of the measured current growth times in the discharge  
with different cathodes allows us to estimate 
the contribution of SEE in the electron avalanche development.

\section{Theoretical model}

\begin{figure}[h!]
\includegraphics[width=0.85\linewidth]{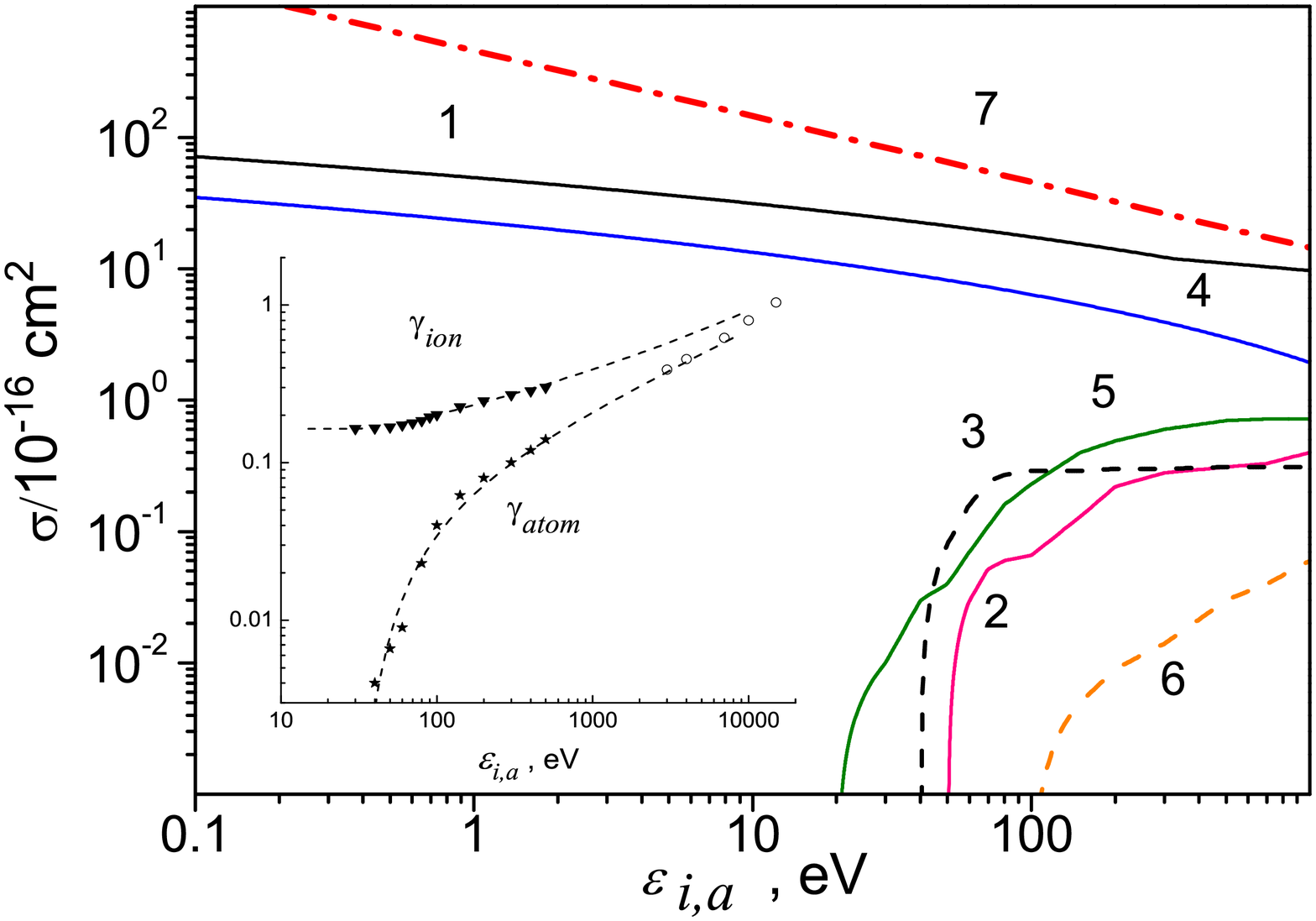}
\caption{Collisional cross sections for ions He$^+$ and energetic atoms 
 He$_f$ in helium: elastic 
 scattering and resonant charge exchange (1), elastic 
 scattering (4), excitation (2,5), 
ionization (3,6), and collisional excitation transfer (7).
 1,2,3 for He$^+$ and 4,5,6,7 for He$_f$.
 Insert: electron emission yield  
$\gamma_i$ for He$^+$ and $\gamma_a$ for He$_f$  
 \cite{Bokhan2007} (triangles, stars),  \cite{Barag} 
 (circles) and dashed curves are analytical fitting.
}
\label{cross}
\end{figure}

For accurate modeling of the HVP discharge for our experimental 
 conditions 
 the energy distribution functions should be calculated for
 charged particles and energetic neutrals.
 Not only electrons, but also ions and energetic atoms gain 
 the energy sufficient for ionization and excitation reactions with 
 background atoms.
 The distribution functions
 for electrons $f^e(t,x,\vec v)$, ions $f^i(t,x,\vec v)$ 
 and energetic atoms $f^a(t,x,\vec v)$
 are calculated with solving the kinetic equations
\begin{equation}  \label{kine}
\frac {\partial f^e}{\partial t}+ \vec v_e\frac {\partial
f^e}{\partial x}
-\frac {e\vec E}{m_e}\frac {\partial f^e}{\partial \vec v_e}=
J_e,\quad n_e=\int f^ed\vec v_e,
\end{equation}
\begin{equation}  \label{kini}
\frac {\partial f^i}{\partial t}+ \vec v_i\frac {\partial
f^i}{\partial x}
+ \frac {e\vec E}{m_i}\frac {\partial f^i}{\partial \vec v_i}
=J_i,\quad
n_i=\int f^id\vec v_i,
\end{equation}
\begin{equation}  \label{kina}
\frac {\partial f^a}{\partial t}+ \vec v_a\frac {\partial
f^a}{\partial x}
=J_a,\quad
n_a=\int f^a\vec v_a,
\end{equation}
where $v_e$, $v_i$, $v_a$, $m_e$, $m_i$, $n_e$, $n_i$, $n_a$ 
 are the velocities, masses
 and densities of electrons, ions and energetic atoms, respectively,
$E$ is the electrical field,
$J_{e,i,a}$ are the collisional integrals.  For electrons
 the integral $J_e$ includes the elastic (momentum transfer) scattering, 
impact excitation and ionization with background atoms
\cite{Alves,Ralch}. 
 After ionization events electrons have the energy and angle
 distributions proposed in \cite{Opal,33}. 
 For ions the integral $J_i$ describes the elastic collision \cite{Cramer},  
 resonant charge exchange, or backward elastic 
 scattering \cite{Cramer}, ion impact 
 excitation \cite{Okasaka} and ionization \cite{Gilb}.
 The scattering of ions 
with the background atoms leads to creation of
the energetic atoms. The  atoms He$_f$ are removed 
from simulations if their energy  becomes smaller than 
1 eV after a collision.
 For energetic atoms, the integral $J_a$ includes 
elastic scattering \cite{Jordan},  impact 
excitation \cite{Kempter},   ionization \cite{Hayden} 
and collisional excitation transfer (CET).
 These CET reactions,  He$^*$+He$_f\rightarrow$He+He$_f^*$, 
 provide a comparable input into the production of 
 excited energetic atoms.
The cross section for 1$^1$S - 2$^1$P excitation transfer $\sigma_r$ 
in helium was calculated in Ref.\cite{Watanabe}. 
In our simulations we took  
              $\sigma_r=4.6 \times 10^{-14} / \sqrt {\varepsilon^a}$, 
cm$^2$,
      where  $\varepsilon^a$ is the energy of energetic atoms 
 measured in eV.
 In Fig. \ref{cross}, the cross sections of collisions 
 with the backgound atoms are shown for  He$^+$ and  He$_f$.

 The Poisson's equation describes the  potential 
 $\phi$ and electrical field  $E$ distributions      
\begin{equation}  \label{Poisson}
\bigtriangleup \phi =4\pi e \left(n_e - n_i \right),
\quad
\vec E=-\frac{d\phi}{d x} \;
\end{equation}
with the following boundary conditions.
At the anode $\phi=U_a$, and  
$U_a$=0, -100~V, -200~V in different variants. 
At the cathode  
\begin{equation}  
\phi=U(t)=U_0\sin(\pi/2\times t/t_p)-j(t)R_{ext}, 
\quad 
 t<t_p;
\end{equation}  
\begin{equation}  
\phi=U(t)=U_0-j(t)R_{ext}, 
\quad 
 t>t_p,
\end{equation}  
where $U_0$ is the voltage amplitude, $t_p$=10~ns is the time of
 increasing $\phi$ from 0 to $U_0$, $j$ is the discharge current,
and $R_{ext}$ is the resistivity of external circuit. 

In simulation we assume that the ions and energetic atoms passing 
the mesh-anode are disappeared with the probability $\delta$=0.3, 
as in the experiment the transparency of the mesh is 0.7. 
As known from the experiment the mesh-anode is under 
floating negative potential, therefore a part of electrons is repelled
 from the mesh-anode.
 In simulations we found that the variation of $U_a$ from 0 to -200~V 
does not influence the breakdown development. 
Electrons, ions and energetic particles approaching the cathode 
 make a contribution to the electron emission and are subtracted
from further simulations.
The Boltzmann equations (1)-(3) are solved with PIC MCC algorithm 
 \cite{birdsall}
   self-consistently with the Poisson's equation (4). 
The time step for electrons is $\Delta t_e=4\times 10^{-16}s $, which is
 much less than the characteristic times of the system,
 $\Delta t_e \ll \Delta x /v_e$ and  $\Delta t_e\ll 1/\omega_p$,
 where $\omega_p$ is the plasma frequency 
and for our discharge conditions 
 $\omega_p \approx $10$^{12}$s$^{-1}$. 

The electron emission is provided by DS photons, electrons, 
ions and energetic atoms fluxes impinging the cathodes surfaces.
 The electron yield for ions, energetic atoms shown 
 in Fig. \ref{cross} (insert).
For simulation of SEE emission we use the data shown in
Fig. \ref{see}. In particular, for CuAlMg alloy the $\gamma_e$ for 
 activated state was used in calculations.

We assume that the resonant photons with a Doppler-shifted frequency 
 reach the cathode instantly without reabsorption in plasma.
These photons are radiated due to atoms excitation
 by ions or energetic atoms impact. Also the DS photons appear from 
 the excited energetic atoms created in CET reactions.
 The coefficient of 
 photoemission of $\gamma_{ph}$=0.3 were taken from 
 Ref. \cite{Bokhan2007} in which the similar discharge was studied.
 More detailed description of our physical
 model can be found in Refs.\cite{Sch2014,Sch2015}.

\section{Results of experimental and theoretical study}

 In our previous study \cite{Sch2015,Sch2016} 
 we have found that the SEE was a major process 
 in electron production in the final stage of the breakdown.
 From the beginning the high energy electrons are accumulated 
 in the discharge volume.  They 
 oscillate  between two powered cathodes through the mesh-anode 
 which is practically transparent for them.
 The characteristic time of the breakdown is much smaller than 
  the electron thermalization time. Therefore since 
  the voltage on the cathodes begins to decrease with 
 the current growth, the high energy electrons overpass 
 the  potential drop over the cathode sheath and reach the cathode.

Now let us consider the effect of enhanced secondary 
 electron emission on the voltage waveform during breakdown.
In the experiment, the waveform  was measured
 in the discharge cells with different cathodes made from 
Ti, SiC and CuAlMg-alloy materials.
In simulations, using different values of $\gamma_e$ 
 corresponding to Ti, SiC and CuAlMg-alloy materials we mimic
 the different cathode cases.

\begin{figure}[h!]
\includegraphics[width=0.85\linewidth]{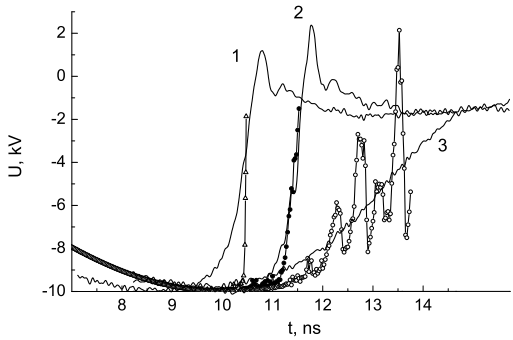}
\caption{Voltage waveforms measured (solid lines) and calculated
 (lines with symbols) for the cathodes made 
from CuAlMg-alloy (1), silicon carbide (2) and titanium (3)
 for $U_0$=10 kV and P=25 Torr.  
}
\label{volt}
\end{figure}

In Fig. \ref{volt}, both
  measured and calculated waveforms with 10~kV-amplitude of  voltage pulse 
on the discharge gap are shown for different types of
 cathodes  for P=25~Torr.
\begin{figure}[h!]
\includegraphics[width=0.85\linewidth]{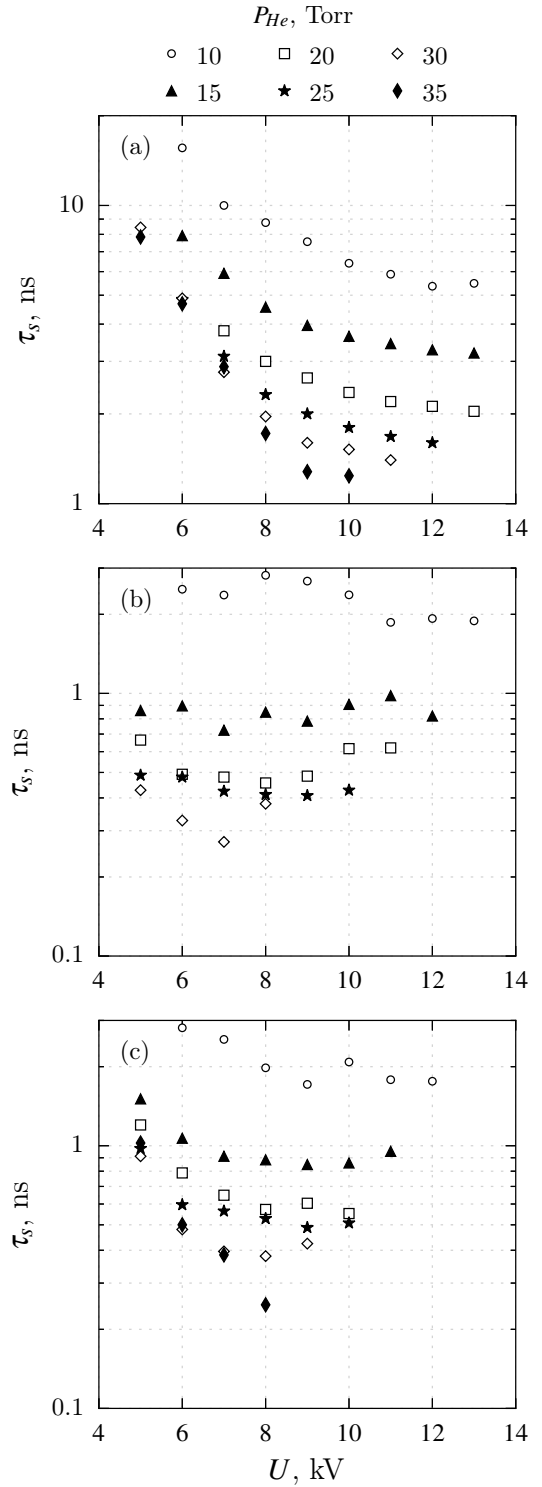}
\caption{Characteristic time of current front rise $\tau_s$ from
pulse voltage amplitude for cathodes made from  Ti (a), SiC (b) and for
 CuAlMg-alloy (c) for different gas pressures.
}
\label{exp_main}
\end{figure}
 It is seen that the breakdown takes place 
 earlier for the larger $\gamma_e$ corresponding to CuAlMg-alloy cathode.
 For the cases of SiC and Ti cathodes the experimental and 
 simulation data well agree showing a steeper
 voltage drop, and consequently a steeper current growth, for larger 
 $\gamma_e$.  The discharge current density increases up to 300~A/cm$^2$ during
 breakdown time $\tau_s$, which is 0.4 ns for SiC case and 2 ns for
  Ti cathode.

 For the case of CuAlMg-alloy cathode the measured $dU/dt$
 differs from the simulated one.
 In the experiment, the cathode material 
 was not in activated state and seems has a smaller $\gamma_e$ of
 the two shown  
 in Fig. \ref{see}. The experimental $dU/dt$ 
 for the SiC and CuAlMg-alloy cases look very similar because 
  their $\gamma_e$ are very close.
 In contrast, in simulation we choose the larger 
  $\gamma_e$ for activated CuAlMg-alloy to show the promising prospects
 of this alloy. As was mentioned above 
 the characteristics 
 of this material is very sensitive to a degree of alloy 
 activation due to magnesium diffusion to the surface 
 and its oxidation by heating. 

 Let us consider the influence of a variation of voltage and gas pressure
 on the breakdown time.
 In Fig. \ref{exp_main}, the switching time $\tau_s$ measured in 
 the experiment
 is shown for different pulse voltages and gas pressures.
 For the cases with Ti and CuAlMg-alloy cathodes shown 
 in Fig. \ref{exp_main} (a) and (c), 
 an increase of  voltage  from 
 5~kV to 7~kV leads to a substantial reduction of $\tau_s$, whereas  
 a variation of $U_0$ from 7~kV to 12~kV only slightly affects $\tau_s$. 
 For the SiC cathode (see Fig. \ref{exp_main} (b)), the switching time
 remains practically constant for the voltage range from 5~kV to 12~kV.
 Only for P=30~Torr, the $\tau_s$ is a more complex function of $U_0$ with 
 a minimum at $U_0$=7~kV.

 The effect of gas pressure on $\tau_s$ is the same for all types of 
 cathodes: with increasing P, $\tau_s$ decreases.
The record switching time for SiC and CuAlMg-alloy is 
$\tau_s<$0.4~ns 
and for Ti-case  the $\tau_s$ is 4-5 times larger.

\begin{figure}[h!]
\includegraphics[width=0.85\linewidth]{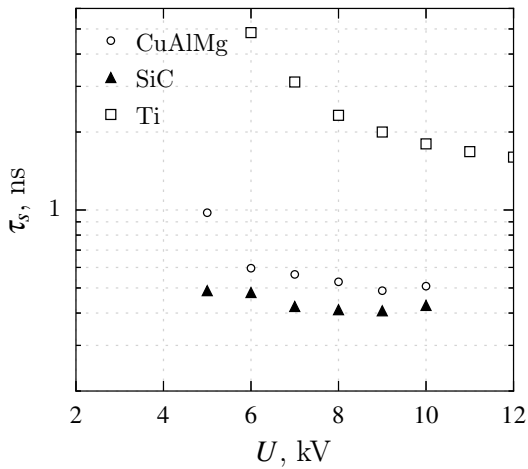}
\caption{Measured $\tau_s$ from
pulse voltage amplitude for cathodes made from  Ti, SiC 
and CuAlMg-alloy for P=25 Torr.
}
\label{exp2}
\end{figure}

\begin{figure}[h!]
\includegraphics[width=0.85\linewidth]{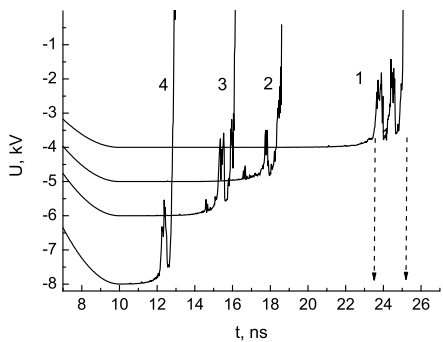}
\caption{Calculated voltage waveform for SiC-cathode for 
 $U_0$=4~kV (1), 5~kV (2), 6~kV and 8~kV,  P=25 Torr.
}
\label{4}
\end{figure}

In Fig. \ref{exp2}, the $\tau_s$-dependence from
pulse voltage amplitude is shown for three types of cathodes 
at P=25 Torr. It is seen that with for the pulse voltage 
amplitude less than 5~kV the $\tau_s$ quickly rises. 
 The calculated voltage profile for various $U_0$ is shown 
 in Fig. \ref{4} for SiC-cathode case. 
 In simulations 
 the voltage begins to increase from t=0 during 10 ns and 
 a delay of breakdown enlarges with a decrease of 
 the pulse voltage amplitude.
 For the 4~kV voltage the waveform differs qualitatively 
 from the variants  with $U_0$=5~kV, 6~kV and 8~kV and 
 reflects the larger switching time.
 
In conclusion, in the experiment and PIC MCC simulations we have 
 studied the 
 influence of different  e-e emission yields, pulse voltages 
  and gas pressures on the  breakdown time
 in high voltage pulse discharge in helium.
 Our previous simulations \cite{Sch2015,Sch2016} have 
 demonstrated that  e-e emission makes the main contribution in 
 the current development on the final stage of breakdown. 
 Therefore we have studied 
 the breakdown  with of three types of cathodes which have very different 
 electron yield. The cathodes made from
 Ti, SiC and  CuAlMg-alloy  were implemented in the experimental 
 sells. The voltage waveforms were measured and calculated for
 the pulse amplitude ranged of 4~kV-12~kV and gas pressure 
 from 10~Torr to 35~Torr. 
  The record switching time $\tau_s \approx $~0.4~ns was
 registered in the experiment and simulations for 
the cases with 
 SiC and  CuAlMg-alloy cathodes, and for titanium cathode with
 smaller e-e yield  the $\tau_s >$2~ns.
  We also found out that the increase of gas pressure
 from 10 Torr to 30 Torr helps to decrease switching time. A decrease
 of pulse 
  amplitude from  10~kV to 5~kV weakly changes $\tau_s$, but
  with further decrease of voltage amplitude ( $U_0<$ 5kV),
  the $\tau_s$ quickly increases.

 Thus there is a specific range of discharge parameters,
 $U_0$=5-10~kV and P=15-35~Torr,
 within that  the record switching time $\tau_s<$1~ns can be achieved.
 For $U_0>$11~kV and P$>$35~Torr the discharge operation transits
 to filament mode. Additionally to these range
 of plasma parameters the cathode material should have
 the secondary electron emission yield not less than
 $\gamma_e$ of the SiC and CuAlMg-alloy materials.

\begin{acknowledgments}
 We are grateful to the Russian Science Foundation 
for supporting this work, project 14-19-00339. The development of software
 was partly supported by by grant of Russian
 Foundation of Basic Research No. 15-02-02536. 
\end{acknowledgments}

\bibliographystyle{IEEEtran}

\begin{thebibliography}{34}
\bibitem{1}
 Mesyats G and Yalandin M 2005 Phys. Usp. {\bf 48(3)} 211 
\bibitem{2}
 Generation of Runaway Electron Beams and 
 X-Rays in High Pressure  Gases 2016 
 {\bf 1,2} Ed. by  Tarasenko V F 
 {New York: Nova Science Publ., Inc.} 


\bibitem{3}
 Bokhan P.A.  
 Pumping of gas lasers by runaway electrons 
generated in an open discharge. Encyclopedia of Low-Temperature 
Plasma. 2005 Ed. by  Fortov V E {\bf XI - 4} 316  (in Russian) 

\bibitem{4}
Khomich V Y and Yamschikov V A 2011 Plasma Physics Reports {\bf 37} 1145 

\bibitem{5}
 Shulepov M, Erofeev M,  Ivanov Y, Oskomov K and Tarasenko V 2015
 J Physical Science and Application {\bf 15} 33 

\bibitem{6}
Bokhan P A, Gugin P P and Zakrevskii D E 2016 Quantum Electron 
 {\bf 46} 782

\bibitem{7}
Efremov A, Koshelev V, Kovalchuk B, Plisko V and  Sukhushin K 2011
 Instrum. Exp. Tech. {\bf 54} 70 

\bibitem{8}
 Carman R J, Kane D M and Ward B K 2010 J. Phys. D: Appl. Phys. {\bf 43} 025205 

\bibitem{9}
Bokhan P A, Gugin P P, Lavrukhin M A and Zakrevsky Dm E 2013 
Phys. Plasmas {\bf 20} 033507 

\bibitem{10} 
Bokhan P, Gugin P, Zakrevsky Dm and Lavrukhin M 2013
 Tech. Phys. Lett.  {\bf 39} 775

\bibitem{Sch2014}    
 Schweigert I V,  Alexandrov A L, Zakrevsky Dm E and  Bokhan P A 
 2014 Phys. Rev. E, {\bf 90} 051101(R) 

\bibitem{Sch2015} 
Schweigert I V, 
 Alexandrov F L, Bokhan P A and  Zakrevsky Dm E 2015  
 Plasma Sources Sci. Technol. {\bf 24} 044005 

\bibitem{Sch2016} 
Schweigert I V, 
 Alexandrov F L, Bokhan P A and  Zakrevsky Dm E 2016   
Plasma Phys. Reports {\bf 42} 666 

\bibitem{exp_d}
  Bokhan P A,  Gugin P P, Lavrukhin M A, Schweigert~I~V,  
Alexandrov  A L and  Zakrevsky Dm E 2016 Switches Based on the Open 
Discharge with Counter-Propagating Electron Beams 
In: Generation of runaway electron beams and x-rays in high pressure 
gases {\bf 1} Techniques and measurements, Ed. by Tarasenko V F 
(New York: Nova Science Publishers Inc) 221


\bibitem{Ti}
Walker C G H, EL-Gomati M M, Assa'd A M D and Zadrazil M 2008
 Scanning {\bf 30} 365
\bibitem{SiC}
 Viel-Inguimbert V 2003 
in {\it Proc. of 28th International Electric Propulsion Conference}, 
Toulouse 1
\bibitem{Lep} 
Lepeshinskaya B N   and Stuchinskiy G B 1960 Fizika Tverdogo Tela 
(Physics of the Solid State), {\bf 2} 1328 (in Russian).



\bibitem{Alves}
Alves L L, Bartschat K, Biagi S F et al 2013 J. Phys. D: Appl. Phys.
{\bf 46(33)} 334002 

\bibitem{Ralch}
Ralchenko Y, Janev R K, Kato T et al 2008 At. Data Nucl. Data Tables
{\bf 94(4)} 603

\bibitem{Opal}
Opal C B, Peterson W K, and  Beaty E C 1971 
J. Chem. Phys. {\bf 55(8)} 4100

\bibitem{33}
Surendra M, Graves D B, and Jellum G M 1990 
Phys. Rev. A {\bf 41(2)} 1112

\bibitem{Cramer}
Cramer W H and  Simons J H 1956 J. Chem. Phys. {\bf 26(5)} 1272

\bibitem{Okasaka}
Okasaka R, Konishi Y, Sato Y and Fukuda K 1987
J. Phys. B: At. Mol. Phys. {\bf 20(15)}  3771


\bibitem{Gilb}
Gilbody H B and Hasted J B 1957  
Proc. R. Soc. A {\bf 240(1222)} 382

\bibitem{Jordan}
Jordan J E and Amdur I 1967 J. Chem. Phys. {\bf 46(1)} 165 

\bibitem{Kempter}
Kempter V, Veith F and Zehnle L 1975 
J. Phys. B: At. Mol. Phys. {\bf 8(7)}  1041

\bibitem{Hayden}
Hayden H C and Utterback N G 1964 Phys. Rev. {\bf 135(6A)} A1575

\bibitem{Watanabe}
Watanabe T 1965  Phys. Rev. {\bf 138(6A)} A1573

\bibitem{birdsall} 
Birdsall C K  and Langdon A B 1985 Plasma Physics Via Computer
Simulation (New York: McGraw-Hill) 

\bibitem{Bokhan2007} 
Bokhan P A,  Zakrevsky Dm E 2007 Tech. Phys. {\bf 52(1)} 104 

\bibitem{Barag}
Baragiola R A, Alonso E V and Oliva Florio A 1979
Phys. Rev. B {\bf 19(1)} 121





\end{thebibliography}

\end{document}